\documentclass[aps, prd, twocolumn, nofootinbib
prd,
]
{revtex4-2}

\usepackage{xcolor}
\usepackage{caption}
\usepackage{subcaption}
\usepackage{amsmath}
\usepackage{amssymb}
\usepackage{graphicx}
\usepackage{dcolumn}
\usepackage{bm}
\usepackage{hyperref}
\usepackage{floatrow}
\usepackage[normalem]{ulem}

\begin{document}

\preprint{APS/123-QED}

\title{Generalized Linear Models of $T_{90}$-$T_{50}$ relation to classify GRBs }

\author{Sourav Dutta}
\email{p22ph008@iitj.ac.in}
\author{Sunanda}
\author{Reetanjali Moharana}
\email{reetanjali@iitj.ac.in}
\author{Manish Kumar}
\affiliation{Department of Physics, Indian Institute of Technology Jodhpur, Karwar 342037, India.}

\date{\today}

\begin{abstract}
Gamma-ray bursts (GRBs) can be classified with their linearly dependent parameters alongside the standard $T_{90}$ distribution. The Generalized linear mixture model(GLM) identifies the number of linear dependencies in a two-parameter space. Classically, GRBs are classified into two classes by the presence of bimodality in the histogram
of T$_{90}$. However, additional classes and sub-classes of GRBs are fascinating topics to explore. In this work, we investigate the GRBs classes in the $ T_{90} {-}T_{50}$ plane using the Generalized Linear Models(GLM) for \textit{Fermi} GBM and  BATSE catalogs. This study shows five linear features for the \textit{Fermi} GBM catalog and four linear features for the BATSE catalog, directing towards the possibility of more than two GRB classes.
\end{abstract}

\maketitle


\section{\label{sec:intro}Introduction}
Gamma-ray bursts (GRBs) can last ten milliseconds to several hundred seconds in the gamma-ray sky. Using the distribution of the period $T_{90}$, refers to the duration at which 90\% of fluence is collected on the detector,  reference \cite{kouveliotou1993identification} noted binary behavior with 222 events in the 1B BATSE catalog. Several attempts are performed with multiple observed GRB parameters like fluence, hardness ratio (higher energy flux compared to lower energies) \cite{fynbo2006no,kann2011afterglows, bromberg2013short, tarnopolski2015analysis, tarnopolski2016analysis} to identify the number of classes in GRBs. All the attempts till now confirm the existence of two classes, notably known as short GRBs (SGRBs), that last a few milliseconds to 2 sec and long GRBs (LGRBs) with $T_{90}>$ 2 sec, significantly. However, the exiting of the third class of GRB has long been debated and is being attempted with the univariate and bivariate analyses of GRBs with Gaussian distribution. For example, \cite{H_1998,refId1} claimed the existence of the third class in BATSE data with less than a $0.5\%$ rejection probability. Analog evidence was found for BeppoSaX \cite{Horv_th_2009} and \textit{Fermi}-GBM GRB catalogs \cite{tarnopolski2015analysis}. Further claims of three class preferences by GRBs are studied with maximum likelihood and $\chi^2$ methods \cite{H_1998,tarnopolski2015analysis}. A detailed study exploring the possibility of a third class of GRBs has been reviewed in \cite{zitouni2015statistical}.

The two classes of GRBs further have different astrophysical origins, hence establishing the classification. LGRBs are associated with  Supernovae (SNe) explosions \cite{galama98,Iwamoto:1998tg}. In contrast, the SGRBs originated during compact star mergers, either neutron star-neutron star (NS-NS) \cite{Eichler:1989ve} or NS-blackhole (BH) mergers due to their observational association with macronovae/kilonovae \cite{Li:1998bw} sGRB:130603B, \cite{Fong2017}, 060614 \cite{Yang:2015pha} and 050709 \cite{Jin:2015dxh}. The advanced  Laser Interferometer Gravitational-Wave Observatory (LIGO) and Virgo gravitational radiation observatories detected the first binary neutron star merger on 17 August 2017. The gravitational wave (GW) known as GW170817 revealed a merger of two low-mass compact objects, whose mass range clearly puts them as neutron stars rather than black hole candidates \cite{LIGOScientific:2017vwq}.

Not long back \cite{10.1093/mnras/stw1835}, used a machine learning algorithm, Gaussian mixed model (GMM) to the $T_{90}$ distribution, for \textit{Fermi} GMB, Beppo, Swift/BAT data and  CGRO)/BATSE to look into GRB classification. Their result confirmed two population components to describe the CGRO/BATSE and Fermi/GBM GRBs in the observer frame and for the Swift/BAT bursts
in the rest frame. Additionally, trimodal distribution explains the Swift GRBs in
the observer frame. However, this result may be affected as it is not complete data. A search for different classes with other observed significant parameters is relevant to establish the robustness of bimodal behavior.

Here, we investigate the classification of GRBs with the GMM algorithm for a single parameter, like T$_{50}$, T$_{90}$, and Hardness-ratio (HR) distributions with larger data sets for Fermi-GBM. Importantly, we investigated the linearly correlated GRB parameters with the Generalized Mixture model (GMM) algorithm. Specifically, the $T_{90}$ vs $T_{50}$ plane collected at the detector, where T$_{50}$ and T$_{90}$ refer to the time to accumulate 50\% and 90\% of the burst fluence, starting at the 25\% and 5\% of fluence level, respectively. From the definition of $T_{90}$ and $T_{50}$, one can expect linear relation between these parameters. Hence, we use the GLM algorithm for \textit{Fermi} GBM and BATSE catalogs to study the correlation coefficient in $T_{90}$-$T_{50}$ plane. We also check for multi-linear relations of $T_{90}$ and $T_{50}$ with the hardness ratio (HR). 

This paper is organized as follows: In section \ref{sec:level2}, we describe the data selection method for \textit{Fermi} GBM and BATSE catalog. Section \ref{sec:level3} describes the GMM method to verify bimodality in the histogram plot of $T_{90}$ and the selection method used for GMM analysis. Section \ref{sec:level4} is dedicated to the GLM method to study the correlation coefficient in $T_{90}$-$T_{50}$,  $T_{90}$-HR and $T_{50}$-HR. Section \ref{sec:level5} discusses the possible outcome of the study.
\section{\label{sec:level2}Data set of GRBs}
We consider two major GRB catalogs for our analysis, $Fermi$ GBM, and BATSE 4B. 
The number of GRBs in the $Fermi$ GBM catalog is 3440 collected from 14$^{th}$ July, 2008 to 19$^{th}$ January, 2023\footnote{\href{https://heasarc.gsfc.nasa.gov/W3Browse/fermi/fermigbrst.html}{https://heasarc.gsfc.nasa.gov/W3Browse/fermi/fermigbrst.html}}. However, the GRB 090626707, have no record for T$_{90}$ and T$_{50}$ values. Hence, we excluded this in the T$_{90}$ distribution study. 2299 valid samples have been considered for investigating the duration and spectral hardness relationship. The BATSE4B current catalog contains 1637 entries (GRBs from 1991 April 21 to 1996 August 14)\footnote{\href{https://heasarc.gsfc.nasa.gov/W3Browse/cgro/batse4b.html}{https://heasarc.gsfc.nasa.gov/W3Browse/cgro/batse4b.html}}. We selected 1233 samples for the study of T$_{90}$ distribution and 1179 samples for investigating the duration and spectral hardness relationship.

We collect the HR data for \textit{Fermi}-GBM  following\cite{10.1093/mnras/stw1835}, where HR is the ratio between the flux from 100 keV to 300 keV to 25 keV to 50 keV, for Comptonised model fit to the spectrum. The HR for BATSE 4B is obtained as the ratio of fluences (flux over the entire burst duration) in the energy range 100-300 keV to 50-100 keV.
 
\section{\label{sec:level3} Analysis Methods}
We first searched for the classification of GRBs in the scattering plot of T$_{50}$ vs T$_{90}$. We first used the Gaussian Mixture model (GMM) following \cite{10.1093/mnras/stw1835}. We then performed another known maximum likelihood classification method, the Generalized Linear Models (GLM) \cite{flexmix} to search for different classes of GRBs we studied GLM on the scattering distribution of T$_{50}$ vs T$_{90}$, T$_{50}$ vs HR, and  T$_{90}$ vs HR. 
\subsection{Gaussian Mixture Model}
We use the machine learning python package scikit-learn{\footnote{\href{https://scikit-learn.org/stable/}{https://scikit-learn.org/stable/}}} to use the GMM method in T$_{90}$ distribution as performed in \cite{10.1093/mnras/stw1835} with at least three times more data. This method helps to avoid the classic histogram method for classifying GRBs. GMM is a mixture of the weighted sum of different Gaussian distributions that describe the number of classes in the data. Assuming k number of Gaussian component $(C_{i},i =1,2....,k)$ the ${i}^{th}$ Gaussian distribution for $x$ number of data is, 

\begin{equation}
N(x|\mu_i,\Sigma_i) =\frac{1}{2\pi}\frac{1}{\sqrt{\Sigma_{i}}} exp\{-\frac{1}{2}(x-\mu_{i})^T\Sigma_{i}^{-1}(x-\mu_{i})\}
\end{equation}

with $\mu_{i}$ mean and $\Sigma_{i}$ covariance. 
The probability distribution function for complete data $X=x_{j}(j =1,2…..,N)$ 

\begin{equation}
    P(X|\omega,\mu,\Sigma) =\sum_{j=1}^{N} \left( \sum_{i=1}^{k} \omega_{i} \;N(x_{j}|\mu_{i},\Sigma )\right),
\end{equation}

where $w_{i}$ is the weight of $i^{th}$ Gaussian distribution.

The GMM method uses an iterative algorithm alternating between \textbf{E}xpectation and \textbf{M}aximization steps (E and M steps) to estimate the parameters of $i^{th}$ number of Gaussian distributions as explained in \cite{10.1093/mnras/stw1835}.
\subsection{Selection Method}
To choose the number of classes, we use the Bayesian information criterion (BIC), also known as the Schwarz\cite{10.2307/2958889} criterion from a finite set of Gaussian distribution models. The model with the smallest value of BIC ($BIC_{min}$) is the preferred model. BIC is defined as:

\begin{equation}
    BIC=p \ln N - 2 \ln P_{max}
\end{equation}

where P$_{max}$, p, N are the maximum likelihood (ML) achieved by the models, the number of parameters of the model, and the sample size, respectively.

To compare the two or more models, BIC difference ($\Delta_i$) is calculated \cite{burnham2004multimodel,tarnopolski2016analysis}. 

\begin{equation}
    \Delta_i=BIC_i-BIC_{min}
\end{equation}

If $0<\Delta_i<2$, then $i^{th}$ model and the minimum BIC model are accepted, and when $2<\Delta_i<6$, $i^{th}$ model becomes a considerable model. For another case where $\Delta_i>6$, indicates strong evidence against the $i^{th}$ model, following $\Delta_i>10$ shows robust proof against it. 
 \subsection{GMM analysis for $T_{90}$ distribution}
 We study the feature of T$_{90}$ distribution of both the \textit{Fermi} GBM and BATSE 4B catalogs using the GMM algorithm.
 \subsubsection{\textit{Fermi} GBM}
\begin{figure}
\begin{subfigure}{.49\textwidth}
\includegraphics[width=\linewidth]{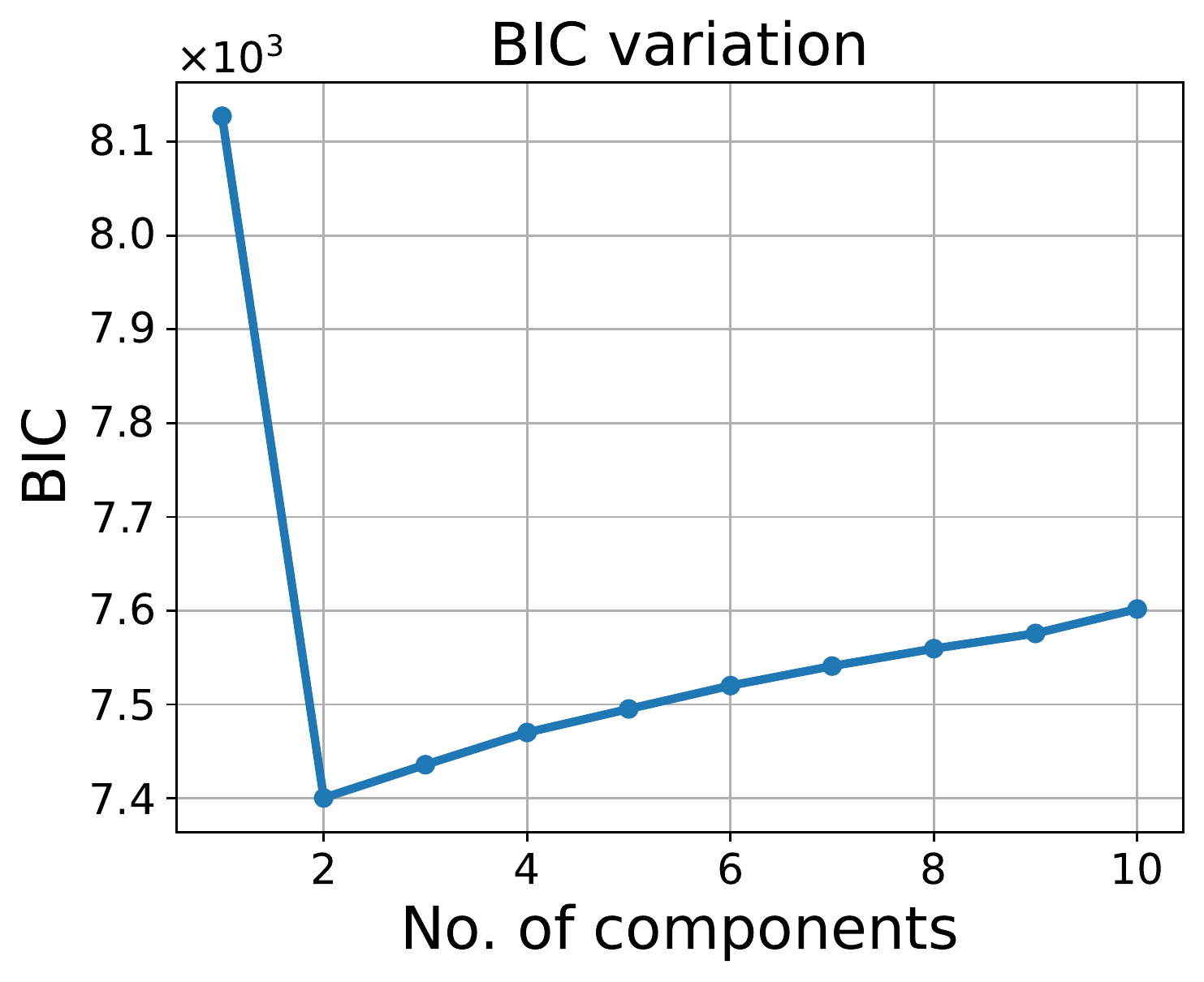} 
\end{subfigure}
\begin{subfigure}{.49\textwidth}
\includegraphics[width=\linewidth]{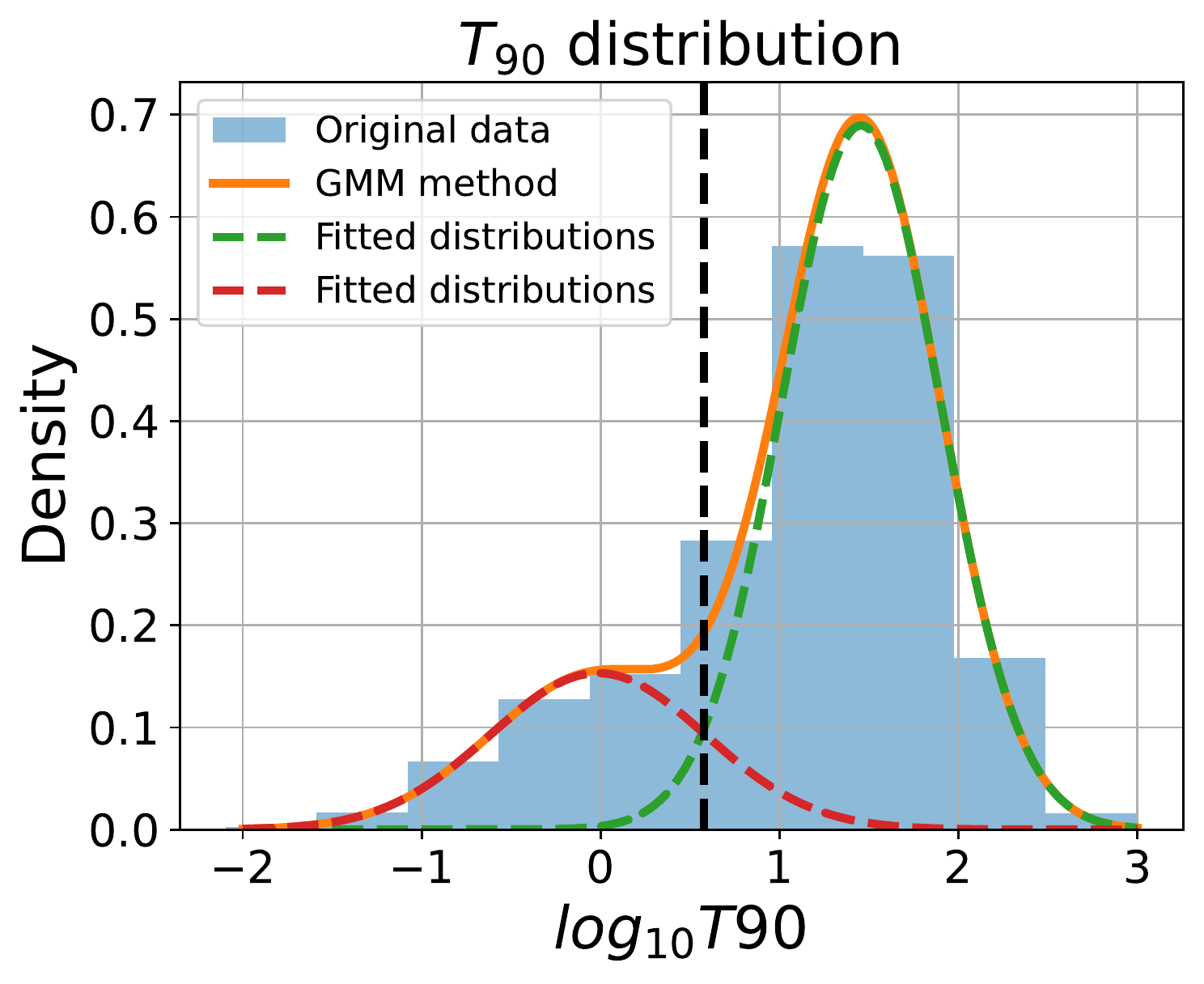}
\end{subfigure}
\caption{GMM analysis for T$_{90}$ distribution of \textit{Fermi} GRB catalog. Left: BIC variations. Right: Distribution plot for two components with the intersection at 3.7 sec.}
\label{fig:fermi_BIC}
\end{figure}

 Fig. \ref{fig:fermi_BIC} shows the variation of BIC values for the number of components and the distribution. This study shows the T$_{90}$ distribution of GRBs fitted with a two-component mixture, namely, Short Gamma Ray Bursts (SGRBs) and Long Gamma-Ray Bursts (LGRBs).

 In the presented dataset, the mean duration of SGRBs is 0.965 sec, with a weight of 0.23 and a standard deviation of 0.6. Similarly, for LGRBs, the mean duration is 28.47 sec, with a weight of 0.77 and a standard deviation of 0.45. The value of BIC$_{min}$ is 7400.4.

 \subsubsection{BATSE 4B}
\begin{figure}
\begin{subfigure}{.49\textwidth}
\includegraphics[width=\linewidth]{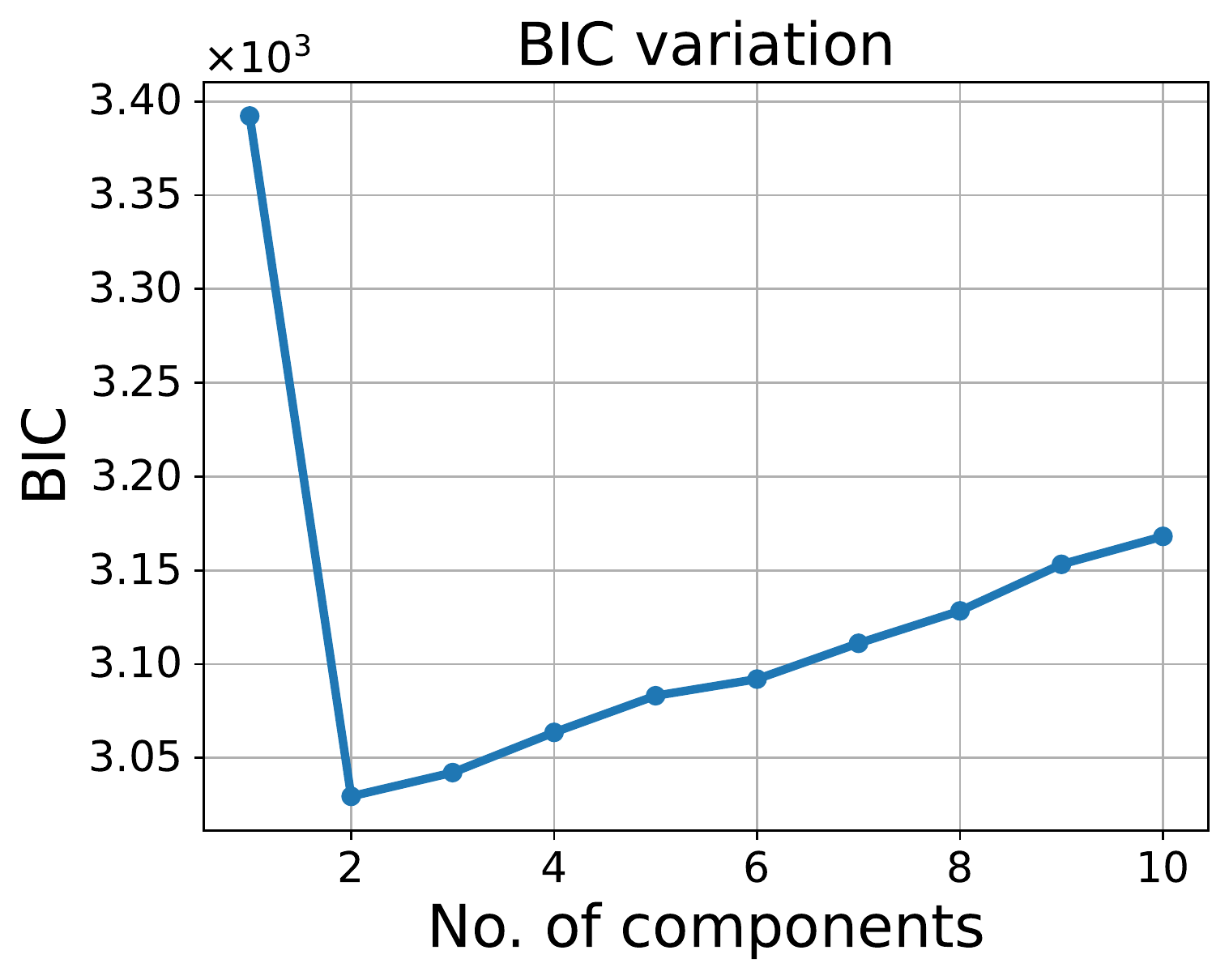}
\end{subfigure}
\begin{subfigure}{.49\textwidth}
\includegraphics[width=\linewidth]{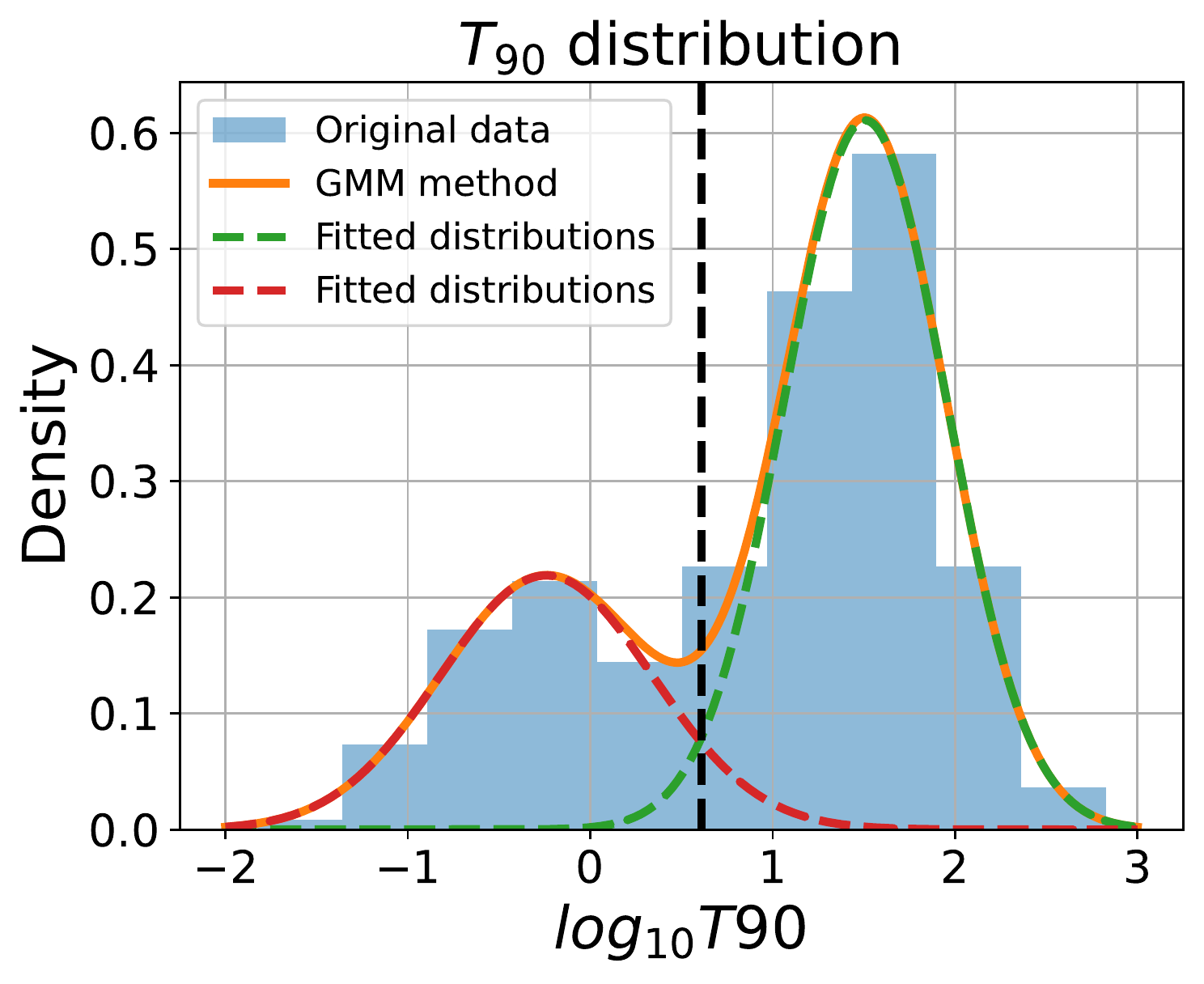}
\end{subfigure}
\caption{GMM analysis for T$_{90}$ distribution of BATSE 4B catalog. Left: BIC variations. Right: Distribution plot for two components with the intersection at 4 sec.}
\label{fig:batse_BIC}
\end{figure}

 Fig. \ref{fig:batse_BIC} shows the variation of BIC values for the number of components and distribution. This study shows the T$_{90}$ distribution of GRBs fitted with the two-component mixture.
 
 In the presented dataset, for SGRBs, the mean duration was 0.57 sec.), with a weight of 0.32 and a standard deviation of 0.58. Similarly, for LGRBs, the mean duration is 32.15 sec, with a weight of 0.68 and a standard deviation of 0.44. The value of BIC$_{min}$ is 1233. 
 
\begin{figure}
\centering
\includegraphics[scale=0.7]{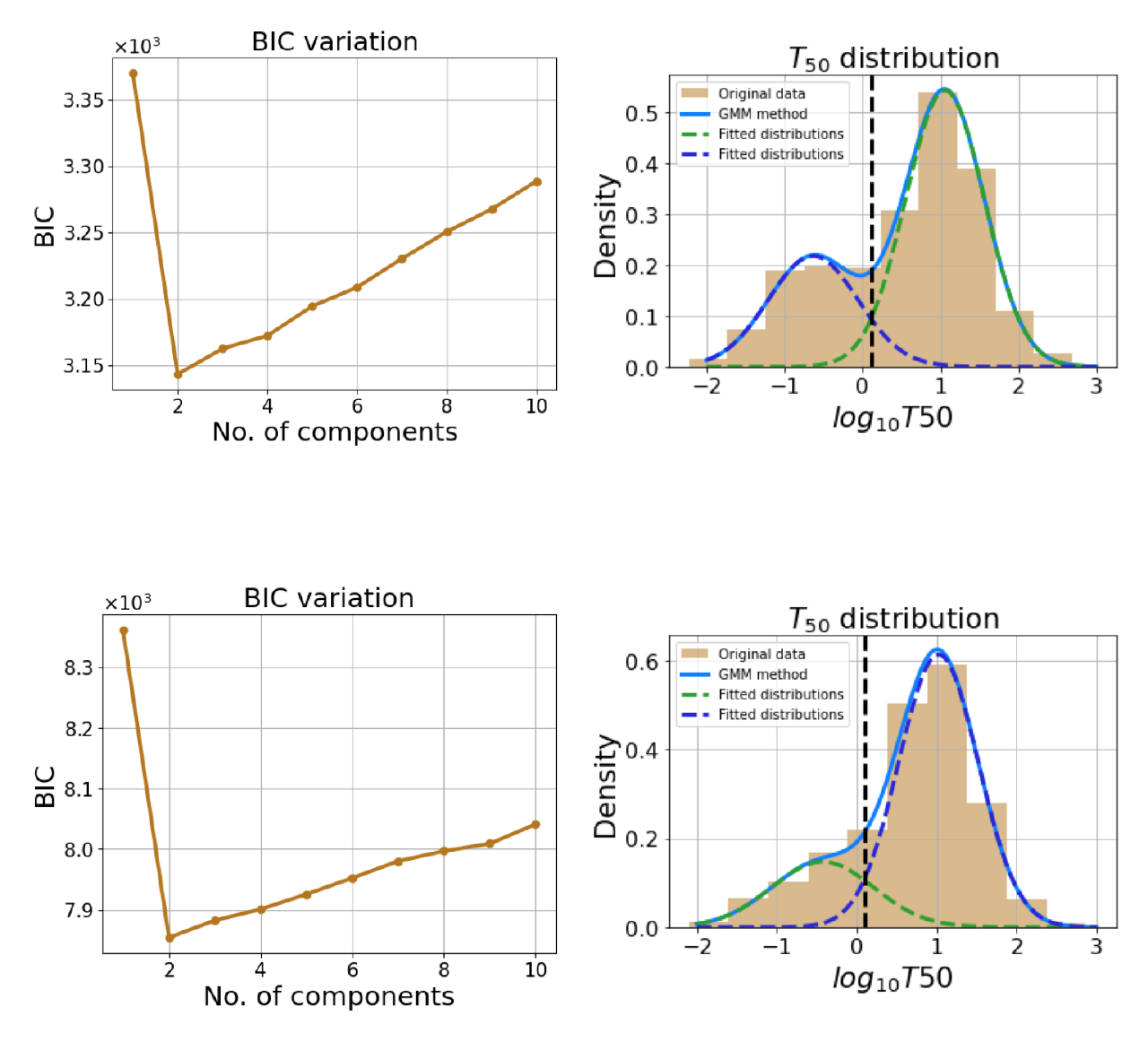}
\caption{GMM analysis for T$_{50}$ distribution for both BATSE 4B (top) and \textit{Fermi}-GBM (bottom) datasets}
\label{fig:t50_combined}
\end{figure}

 The above study has been performed, with nearly one-third of data by other groups for the T$_{90}$ distribution, and found the existence of two GRB classes\cite{10.1093/mnras/stw1835}. We also studied \ref{fig:t50_combined} the BIC variation of T$_{50}$ distribution for both the BATSE 4B and \textit{Fermi}-GBM datasets. The mean distribution of SGRBs is 0.23 sec, with a weight of 0.31 for the BATSE 4B dataset, and 0.37 sec, with a weight of 0.23 for \textit{Fermi}-GBM dataset, while that of LGRBs is 11.41 sec, with a weight of 0.68 for BATSE 4B dataset, and 10.53 sec, with a weight of 0.76 for \textit{Fermi}-GBM dataset. This study again supports two classes of GRBs.
 
 This classification should be validated with the co-dependence of two parameters. Rather than looking for clusters in the scattering plots of two parameters, we checked for the linear dependency of the parameters.  One would expect a linear relation between T$_{50}$ and T$_{90}$, T$_{90}$-HR and T$_{50}$-HR. Hence we use the generalized linear regression mixture (GLM) testing for these parameter spaces to check the number of distinct linear features present.

\section{\label{sec:level4}Generalized Linear Model}
Generalized Linear Models (GLM) are powerful, decisive tools that offer flexible and computationally attractive models for large datasets. Model choice is made important due to this resulting flexibility and complexity. GLM has been used in a wide range of areas from longitudinal data analysis\cite{laird1982random} to penalized spline smoothing\cite{ruppert2003semiparametric} and functional data analysis\cite{Di_2009}. Considering finite mixture models with k components of the form:

\begin{eqnarray}
     h(y|x,\psi)=\sum_{i=1}^{k}\pi_if(y|x,\theta_i) \nonumber \\
    \pi_i\geq0 , \sum_{i=1}^{k}\pi_i=1,
\end{eqnarray}

where y is a dependent variable with conditional density h, x is a vector of independent variables, $\pi_i$ is the prior probability of component i, $\theta_i$ is the component-specific parameter vector of the density function f, and $\psi=(\pi_1,...,\pi_k,\theta_1,...,\theta_k)$ is a vector for all the parameters. The posterior probability for the class j with the 
 observation (x,y) is given by,
 
\begin{equation} \label{GLM_p}
    P(j|x,y,\psi)=\frac{\pi_jf(y|x,\theta_j)}{\sum_{k}\pi_kf(y|x,\theta_i)}.
\end{equation}

The posterior probabilities are used for data classification by assigning each observation to that class with the maximum value of posterior probability. The log-likelihood of a sample of N observations ${(x_1,y_1),...,(x_N,y_N)}$ is given by 

\begin{eqnarray}
    \log L &=&\sum_{n=1}^{N}\log h(y_n|x_n,\psi) \nonumber \\
    &=&\sum_{n=1}^{N}\sum_{i=1}^{k}\pi_if(y_n|x_n,\theta_i).
\end{eqnarray}

This value can not be maximized directly. The most popular method for maximum likelihood estimation for the parameter vector $\psi$ is through the iterative EM algorithm. 

\textbf{Estimate(E)} the posterior class probabilities corresponding to each observation and derivation of the prior class probabilities,

\begin{equation}
\widehat{\pi_i}=\frac{\sum_{n=1}^{N}P(i|x_n,y_n,\psi)}{N}.
\end{equation}

Equation \ref{GLM_p} is used to calculate the posterior probabilities.

\textbf{Maximize(M)} the log-likelihood for each component separately using the posterior probabilities as weights

\begin{equation}
    max\sum_{n=1}^{N}\widehat{p_{ni}}\log f(y_n|x_n,\theta_i),
\end{equation}

where, $\widehat{p_{ni}}$ is given by equation \ref{p_ni}.

\begin{equation}
\label{p_ni}
\widehat{p_{ni}}=\sum_{n=1}^{N}P(i|x_n,y_n,\psi)
\end{equation}

The E and M steps are repeated until the likelihood improvement falls under the pre-specified threshold.
\subsection{GLM analysis}
We have used the "flexmix" model, available in R \cite{flexmix}, for GLM analysis. The "flexmix" model is a general framework for finite mixtures of regression models that implements the EM algorithm. The E-step and all data handling are provided, while the user can supply the M-step to define new models easily. We have used the default generalized linear models (glm) method for clustering the following parameter space. The randomness arises in the BIC values due to random assignment to classes with probabilities $\widehat{p_{ni}}$. To control this randomness, the determination of BIC$_{min}$ was performed $10^5$ times for each parameter space. The results of the same are shown in fig. \ref{fig:BIC_combined} and tabulated in tables \ref{tab:BIC_fermi}.

\subsubsection{T$_{90}$-T$_{50}$}
\begin{figure}
\begin{subfigure}{.49\textwidth}
\includegraphics[width=\linewidth]{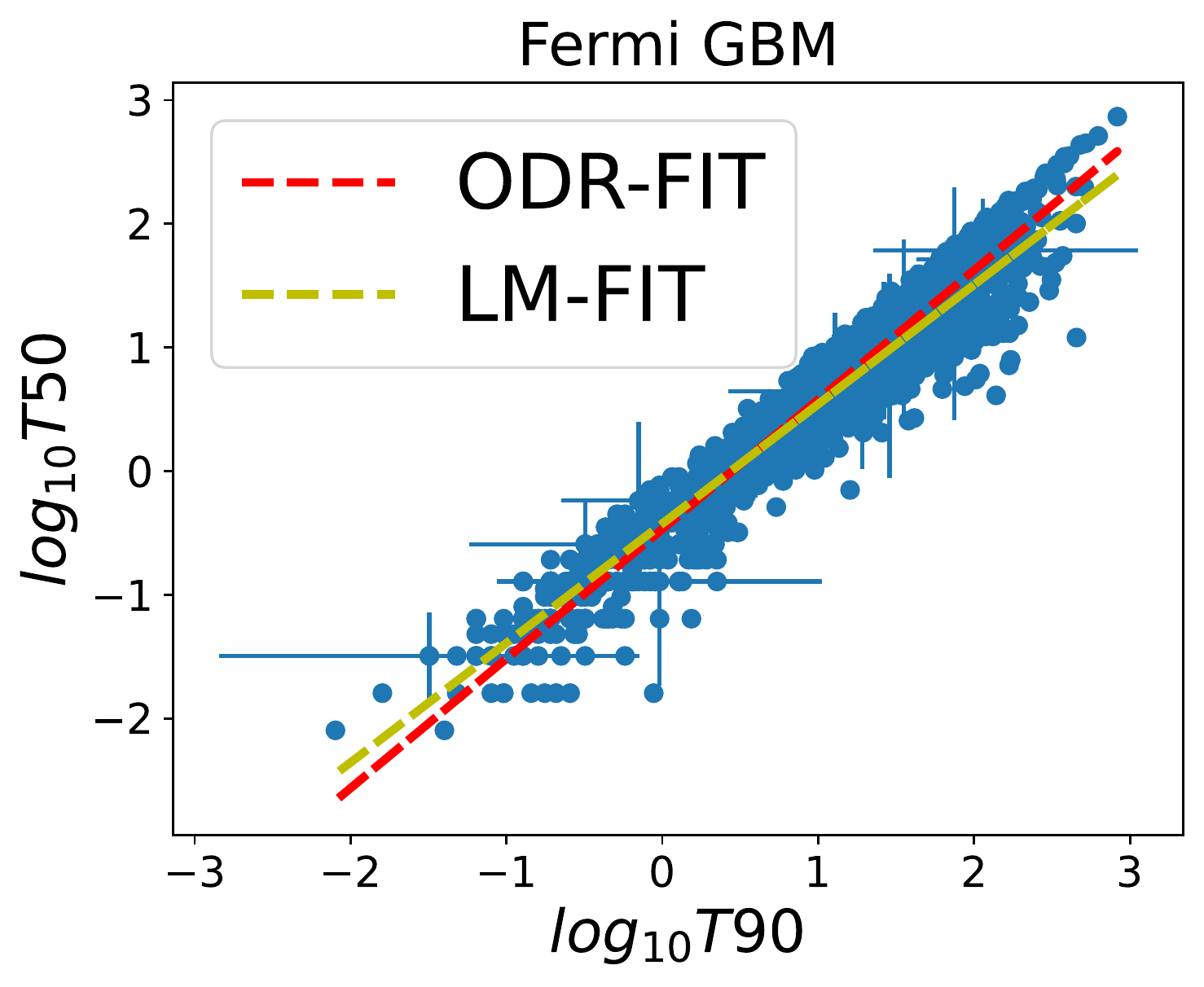}
\end{subfigure}
\begin{subfigure}{.49\textwidth}
\includegraphics[width=\linewidth]{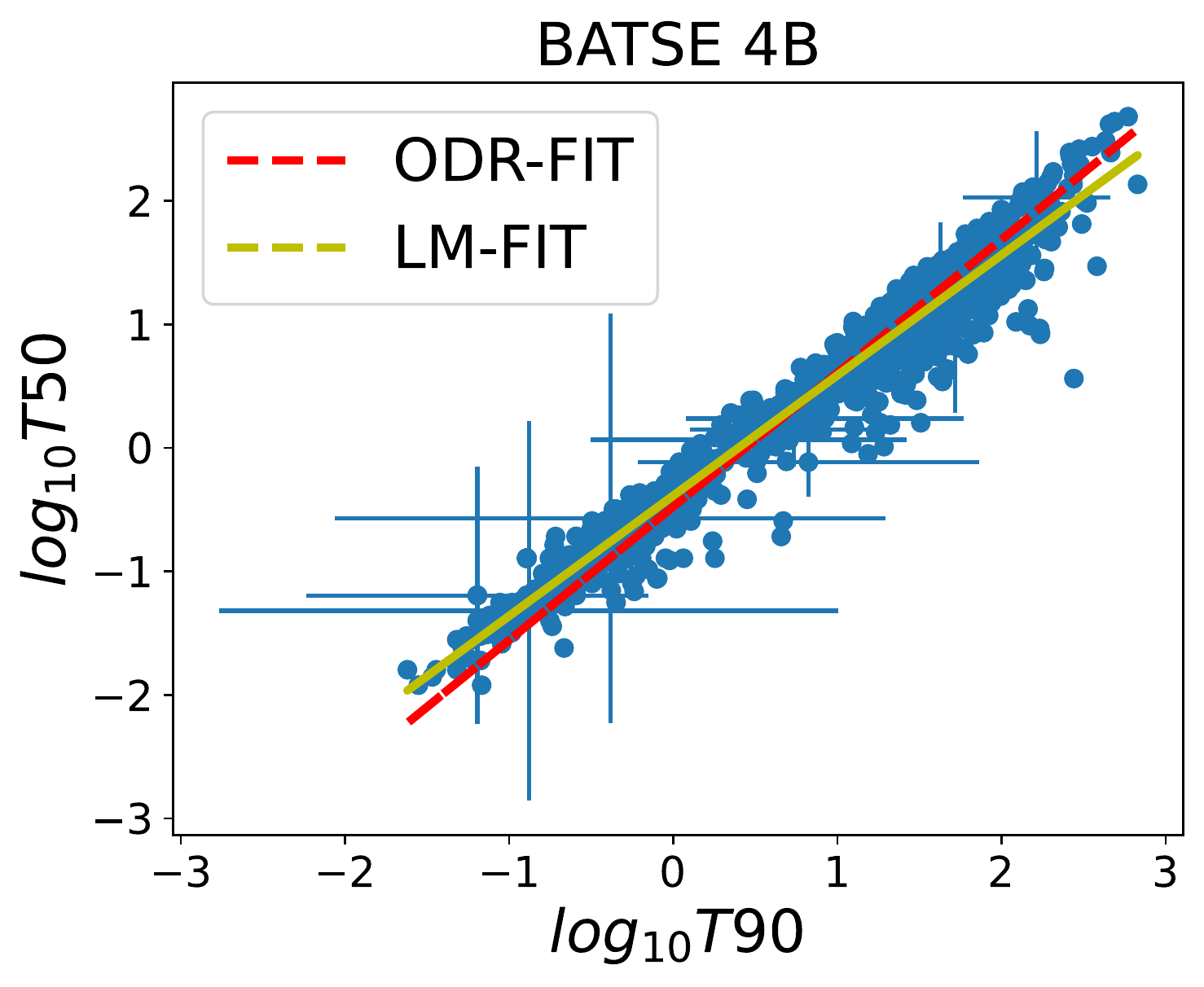}
\end{subfigure}
\caption{$\text{T}_{50} - \text{T}_{90}$ correlation with linear regression.}
\label{fig:LR}
\end{figure}

The correlation coefficient, $\rho$ from linear regression (lm) for T$_{90}$-T$_{50}$ parameter space is 0.96 for the Fermi-GBM dataset. The error bars for T$_{90}$ and T$_{50}$ are significantly large. Hence we also used orthogonal distance regression (ODR) methods to calculate the slope of the linear line, $m_{ODR}$, and found its value 1.05. Figure \ref{fig:LR} shows the results. However, with this analysis, we can not tell the possibility of the existence of multilinear dependence of the parameters. 

\begin{figure}[h!]
\begin{subfigure}{.49\textwidth}
\includegraphics[width=\linewidth]{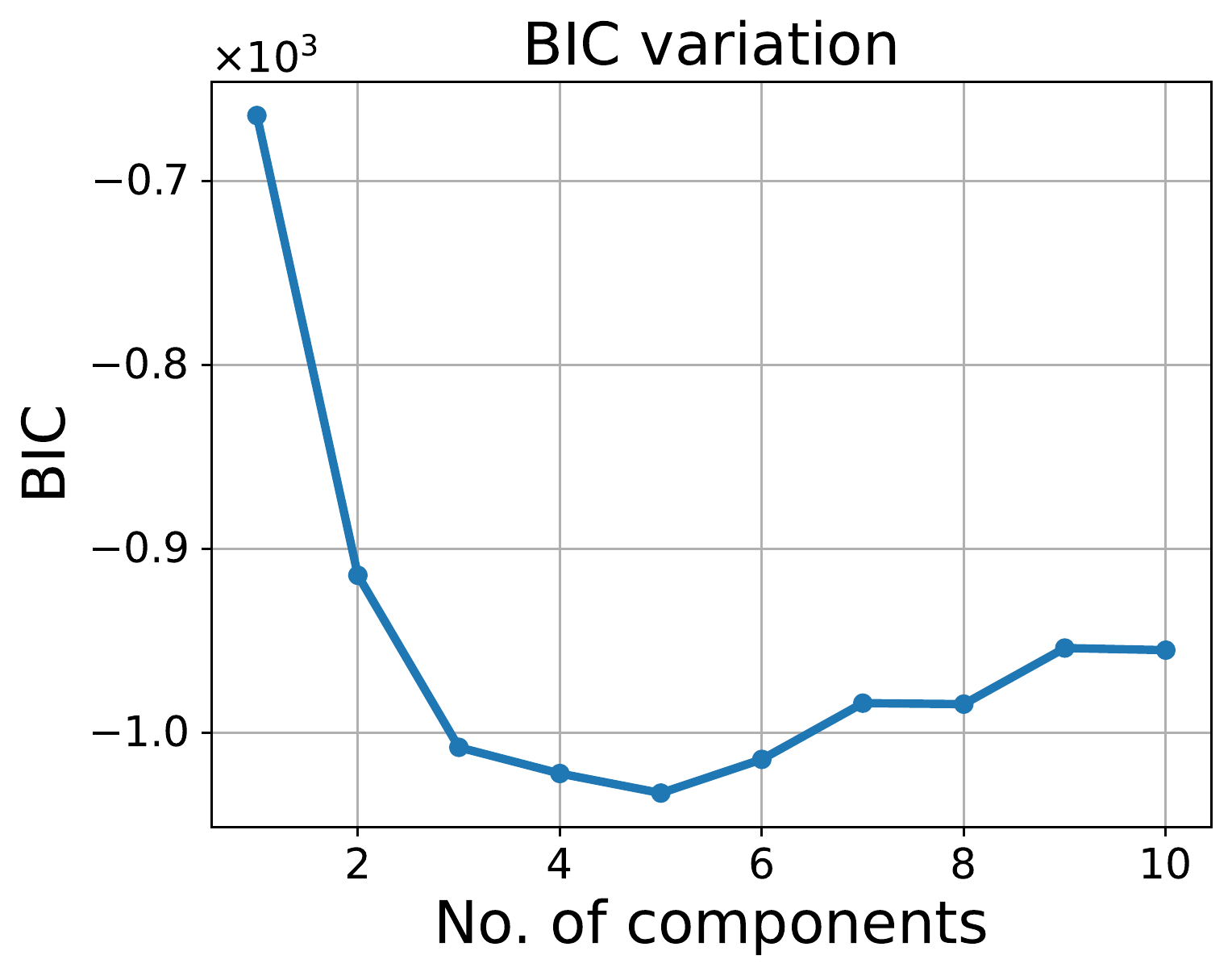}
\end{subfigure}
\begin{subfigure}{.49\textwidth}
\includegraphics[width=\linewidth]{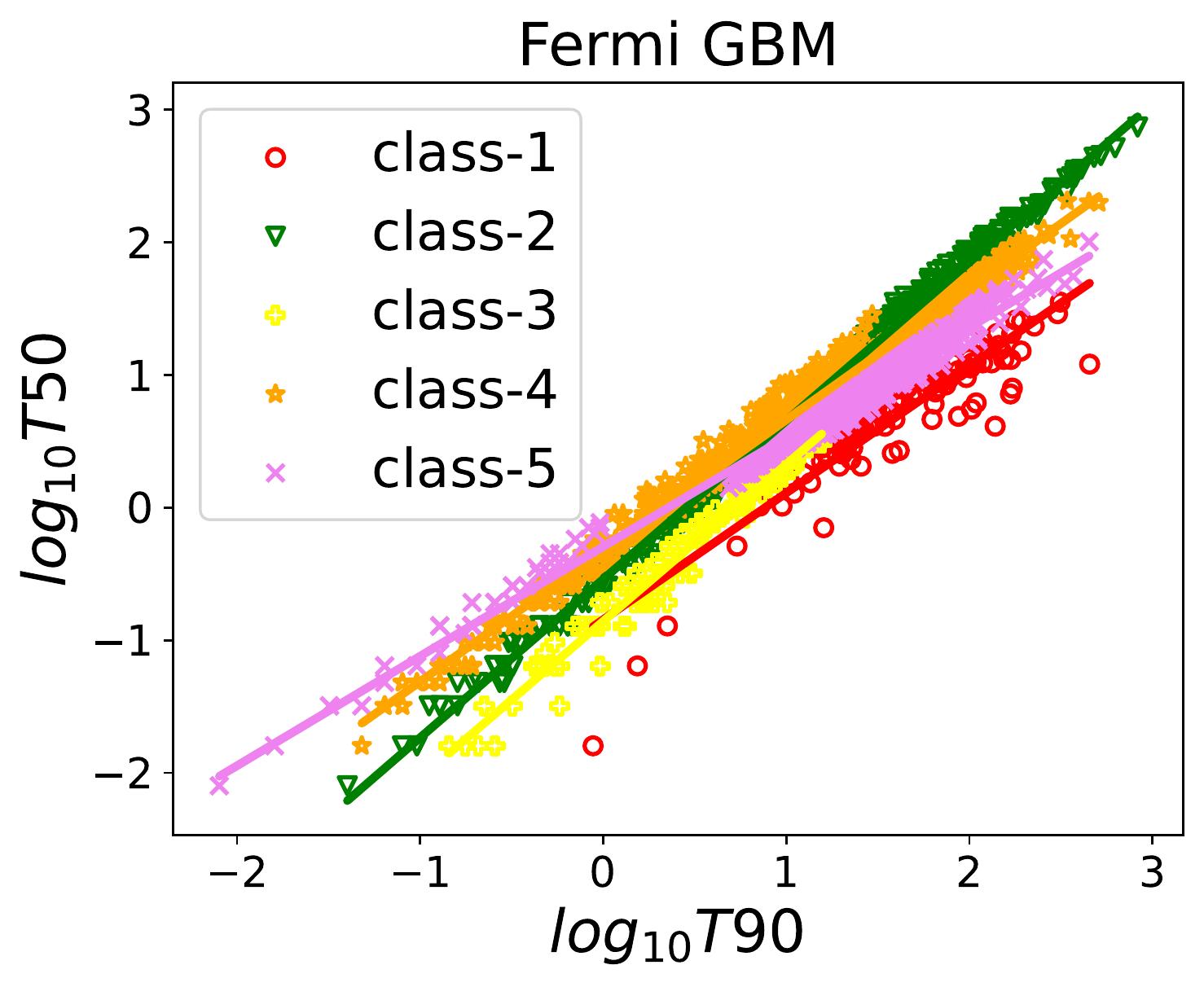}
\end{subfigure}
\caption{GLM analysis for T$_{90}$ vs T$_{50}$ distribution of \textit{Fermi} GRB catalog. Left: Sample plot for BIC variation for different components. Right: The five classes with their linear fit slopes as, 0.95,1.2,1.18,0.99 and 0.82}
\label{fig:GLM_t50_t90_fermi}
\end{figure}

Our analysis through the GLM is shown in fig. \ref{fig:GLM_t50_t90_fermi} for one run. It shows the variation of BIC values for the number of components of the T$_{90}$-T$_{50}$ parameter space corresponding to \textit{Fermi} GBM dataset for one case. In this case, we find $\Delta_4 = 10.8$. Hence the total number of classes is five. On the right side of the fig \ref{fig:GLM_t50_t90_fermi} we show the five groups along with their fitted linear lines for this case. 

\begin{figure}[h!]
	\centering
	\begin{subfigure}{.485\columnwidth}
		\centering
		\includegraphics[width=\textwidth]{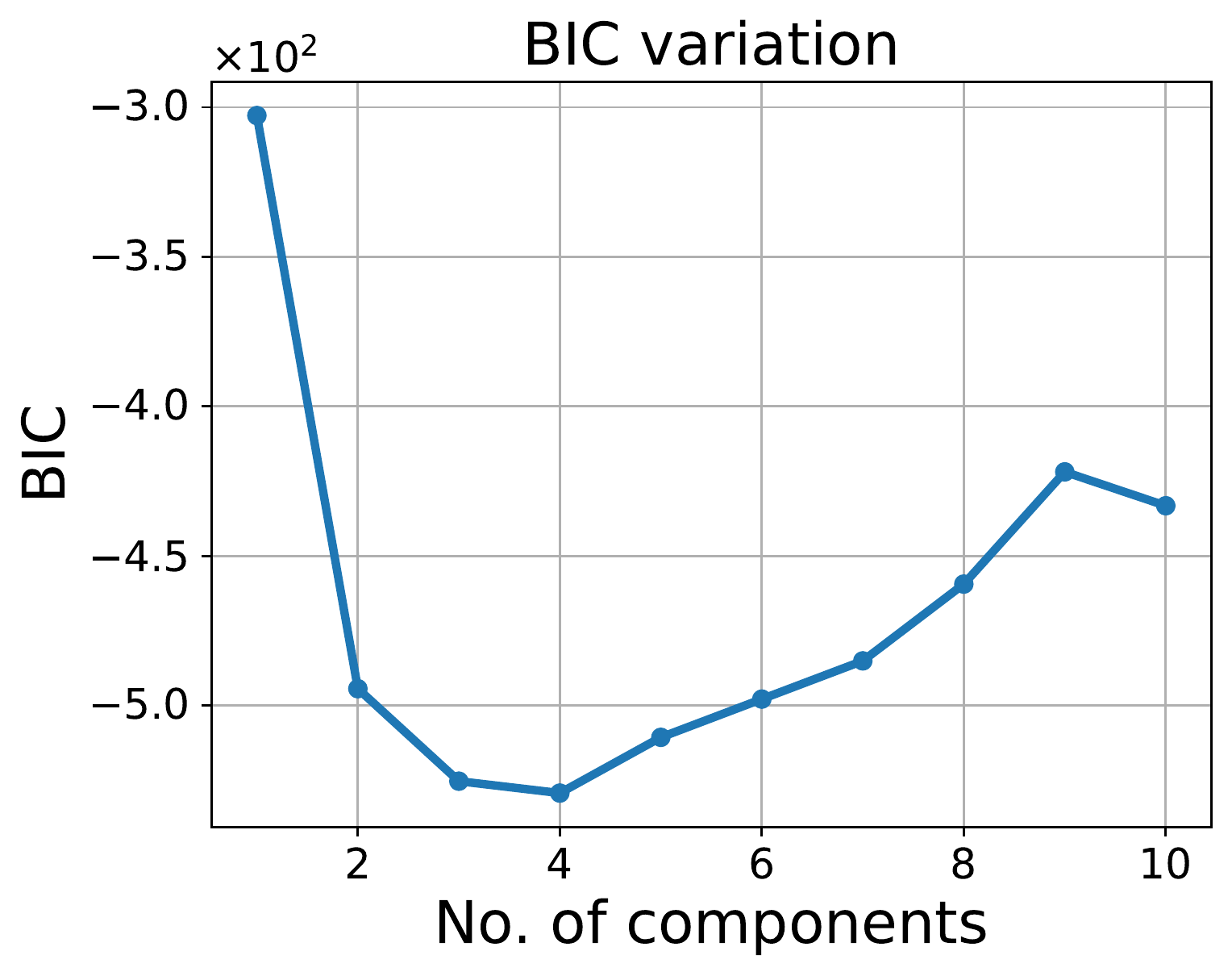}
	\end{subfigure}
	\begin{subfigure}{.485\columnwidth}
		\centering
		\includegraphics[width=\textwidth]{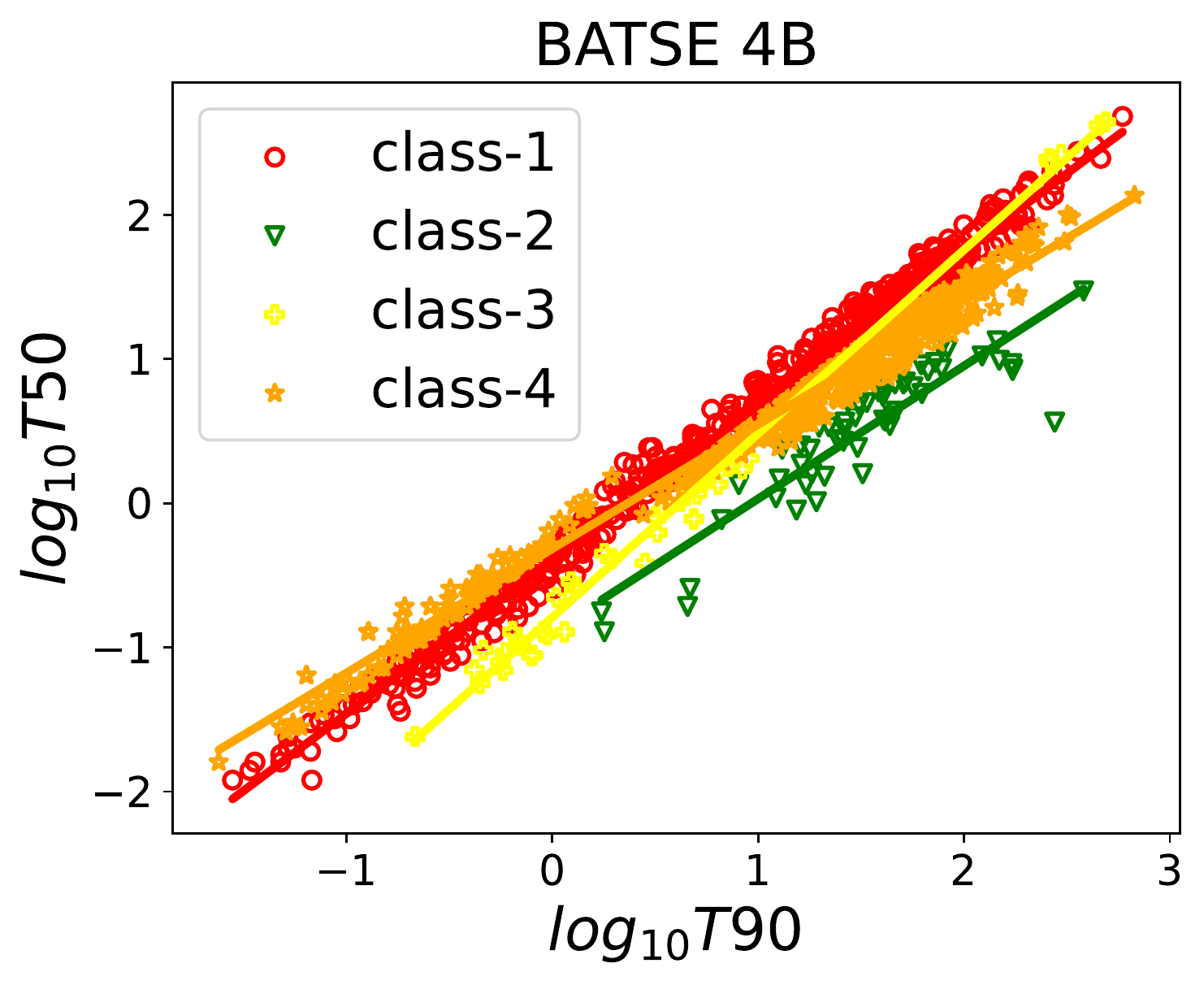}
	\end{subfigure}
	\caption{GLM analysis for T$_{90}$ - T$_{50}$ distribution of BATSE 4B catalog. Left: Sample plot for BIC variation for different components. Right: The four groups with linear slopes, 1.07,0.92,1.3 and 0.8}
	\label{fig:GLM_t50_t90_batse}
\end{figure}

The GLM study for the T$_{90}$-T$_{50}$ parameter space with the BATSE 4B dataset is shown in the figure. \ref{fig:GLM_t50_t90_batse}. In this case $\rho$, is 0.97. In this case, we find $\Delta_3 = 3.97$. Hence the total number of classes is four. We also used orthogonal distance regression (ODR) methods to calculate the slope of the linear line, $m_{ODR}$, and found its value of 1.08.

\begin{figure}[h!]
\centering
 \begin{minipage}{0.48\columnwidth}
\centering
\includegraphics[width=\textwidth]{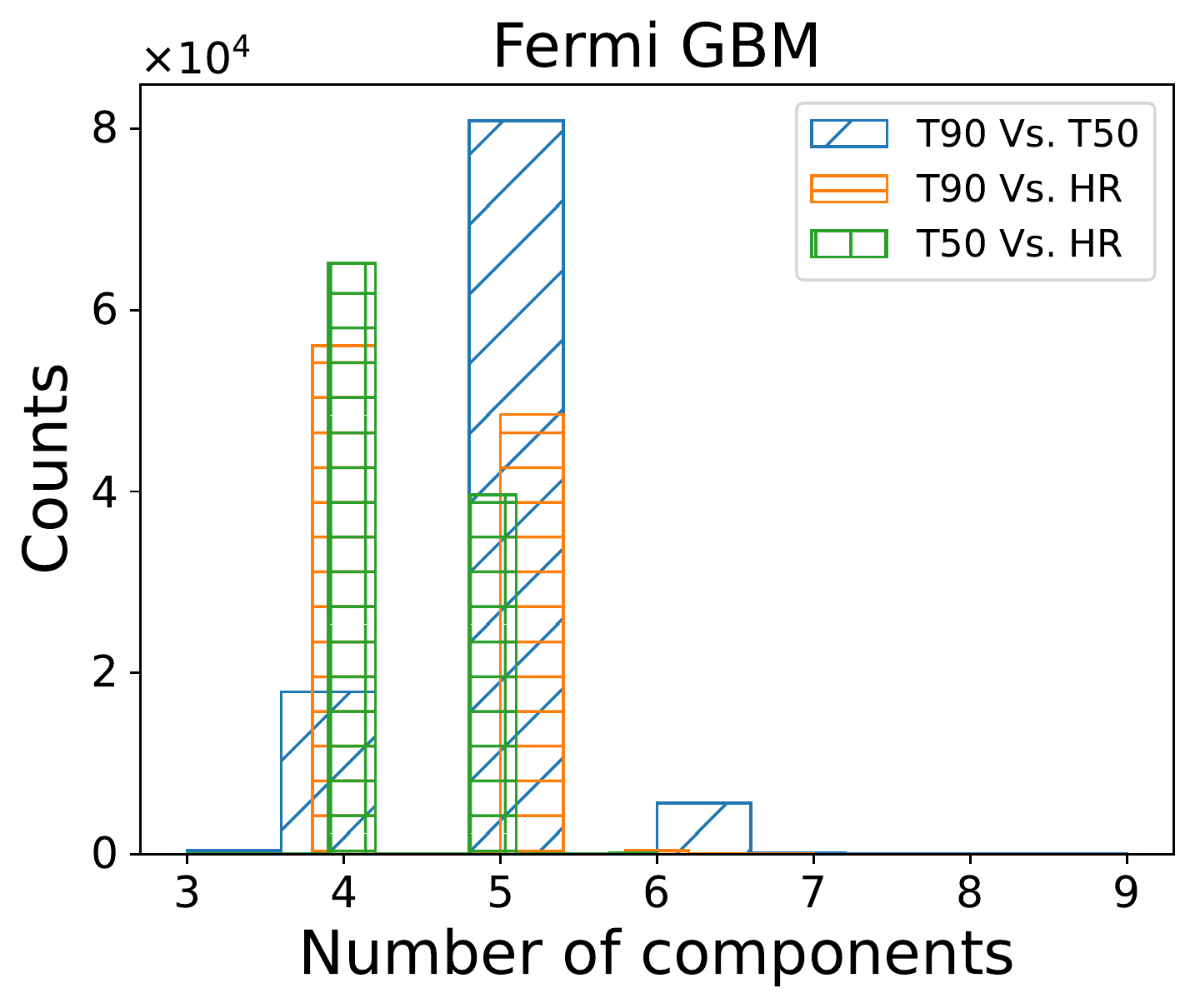}
\end{minipage}
\begin{minipage}{0.48\columnwidth}
\centering
\includegraphics[width=\textwidth]{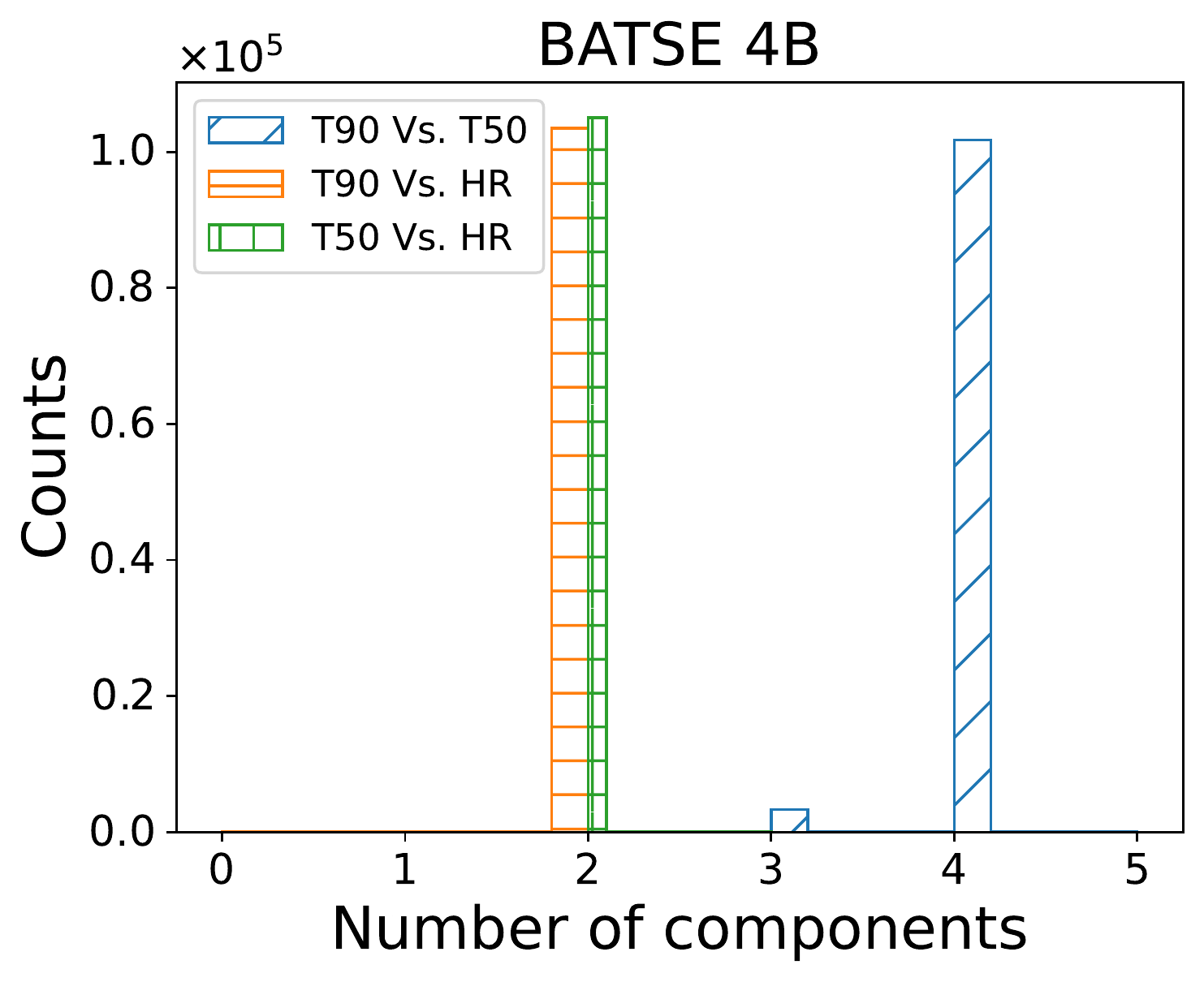}
\end{minipage}
\caption{Distribution of BIC$_{min}$ values for each parameter space}
    \label{fig:BIC_combined}
\end{figure}

To check the randomness in the BIC$_{min}$, we performed the GLM analysis for $10^5$ times for T$_{90}$-T$_{50}$, shown in fig \ref{fig:BIC_combined} with slant strike histogram. We list the details of the occurrence of different classes for both Fermi-GBM and BATSE 4B data in table \ref{tab:BIC_fermi}.

\begin{table}[h!]
    \begin{ruledtabular}
    \caption{Distribution of component numbers for \textit{Fermi}-GBM and BATSE 4B GRB events}
    \begin{tabular}{c c c c}
    x-axis & y-axis & Number of components & \% of occurrence\\
    \hline
    \multicolumn{4}{c}{$Fermi$-GBM}\\
    \hline
    T$_{90}$ & T$_{50}$ & 4 & 17.05\\
    & & \textbf{5} & \textbf{77.00}\\
    & & 6 & 5.35\\
    T$_{90}$ &  HR & 3 &0.02\\
    & & \textbf{4} & \textbf{53.38}\\
    & & 5 & 46.20\\
    T$_{50}$ & HR & 3 & 0.04\\
    & & \textbf{4} & \textbf{62.07}\\
    & & 6 & 0.15\\
    \hline
     \multicolumn{4}{c}{BATSE 4B}\\
    \hline
    T$_{90}$ & T$_{50}$ & 3 & 3.12\\
    & & \textbf{4} & \textbf{96.88}\\
    T$_{90}$ &  HR & \textbf{2} & \textbf{100.00}\\
    T$_{50}$ & HR & \textbf{2} & \textbf{100.00}\\
    \end{tabular}
    \label{tab:BIC_fermi}
    \end{ruledtabular}
\end{table}

\subsubsection{T$_{50}$-HR \& T$_{90}$-HR}
\begin{figure}
    \centering
    \includegraphics[width=\textwidth]{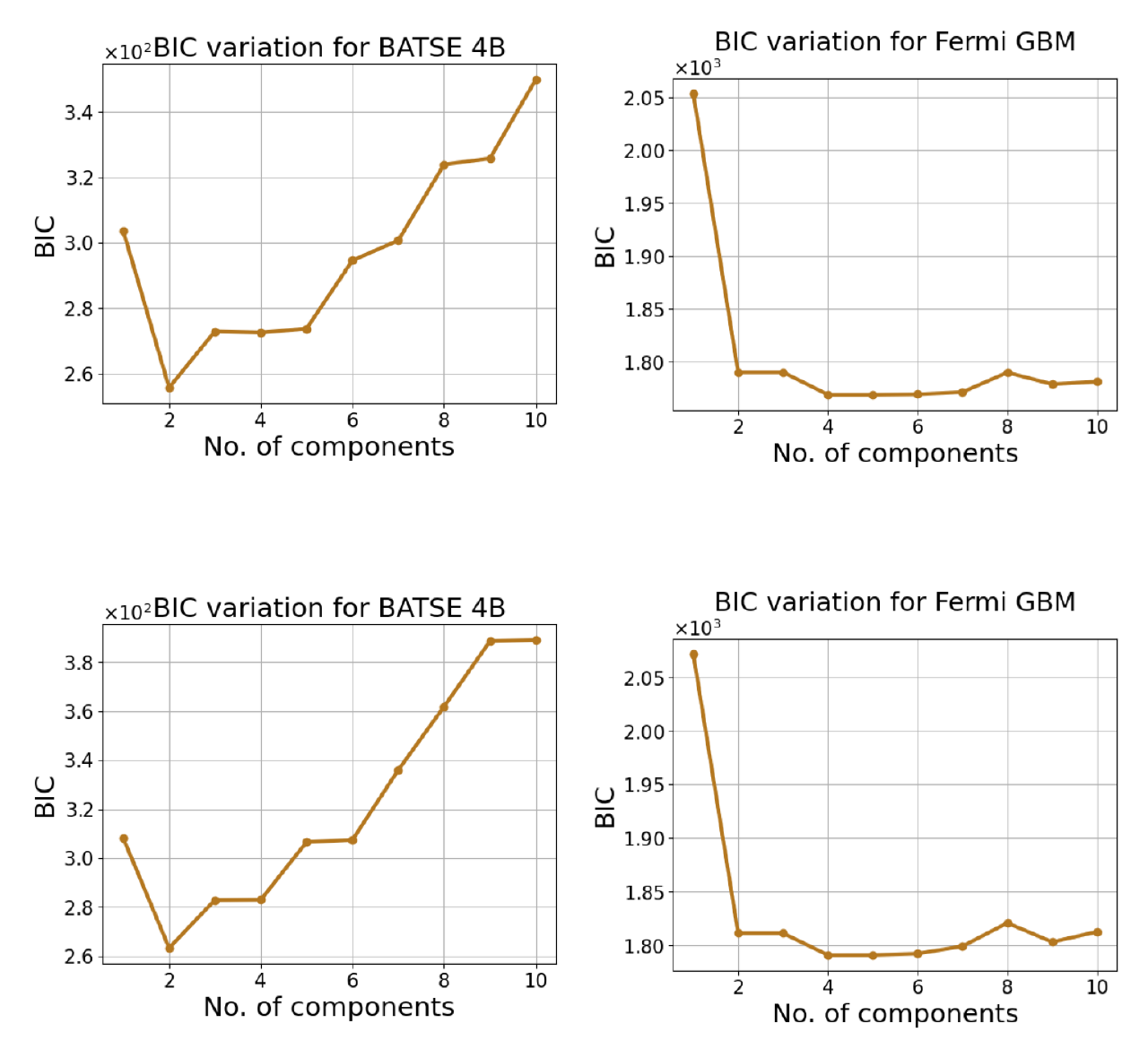}
    \caption{Sample plot for BIC variation for different components for T$_{90}$ vs HR (top) and T$_{50}$ vs HR (bottom) distributions}
    \label{fig:GLM_t50_hr_fermi}
\end{figure}

We study the linear dependence of T$_{90}$-HR \& T$_{50}$-HR to further check the different GRB classes. The correlation coefficient, $\rho$ for T$_{90}$-HR parameter space is -0.32 for the \textit{Fermi} GBM dataset. Our analysis through the GLM is shown in fig. \ref{fig:GLM_t50_hr_fermi}. It shows the variation of BIC values for the number of components of the T$_{90}$-HR parameter space corresponding to \textit{Fermi} GBM dataset for one case suggesting the possibility of more than 3 linear fits to the data. In contrast, the T$_{90}$-HR BATSE 4B dataset suggests bilinear dependency.  

The correlation coefficient,$\rho$ for T$_{50}$-HR parameter space is -0.31 for the \textit{Fermi} GBM dataset. Our analysis through the GLM is shown in fig. \ref{fig:GLM_t50_hr_fermi} right-bottom. It shows the variation of BIC values for the number of components of the T$_{50}$-HR parameter space corresponding to \textit{Fermi} GBM dataset for one case. In this case, we find $\Delta_5 = 0.7$, suggesting the possibility of four different linear dependencies. 
The GLM study for T$_{50}$-HR parameter space with BATSE 4B dataset is shown in fig. \ref{fig:GLM_t50_hr_fermi} left-bottom. In this case $\rho$, is -0.38. In this case, we find $\Delta_3 = 19.6$, again suggesting bilinear behavior. 

\section{\label{sec:level5}Discussion \& Outlook}
The binary classification, long and short, of GRBs is well established. However, the possibility of further classes and sub-classes is still there. Several attempts have been made to find the third class of GRB using different statistical methods. Following \cite{10.1093/mnras/stw1835}, we study the classification with GMM technique using EM-algorithm not only for T$_{90}$ distribution but also for T$_{50}$ distribution. This study showed bimodal behavior in both distributions, even with a larger data set than in \cite{10.1093/mnras/stw1835}. However, the bimodal result can be further established by studying the dependency of different parameters on each other. Motivated by this, we have applied the Generalized Linear Models (GLM) to check the linear dependency of T$_{90}$-T$_{50}$, T$_{90}$-HR, and T$_{50}$-HR for both the \textit{Fermi} GBM and BATSE 4B GRB events. This study indicates 
 the possibility of more than two classes of GRBs, specifically in the linear relation of T$_{90}$-T$_{50}$. Importantly the technique can group the GRB events with a slower increase of $T_{50}$ than $T_{90}$.

\begin{acknowledgments}
The authors acknowledge Science and Engineering Research Board (SERB) Grant No. SRG/2020/001932.
\end{acknowledgments}

\bibliography{glm_grb}

\end{document}